\documentclass{PoS}

\title{Critical Point and Deconfinement from Dyson-Schwinger Equations}

\ShortTitle{Critical Point and Deconfinement from Dyson-Schwinger Equations}

\author{
\speaker{Jan Luecker}$^a$,
Christian~S.~Fischer,$^a$
Leonard~Fister$^b$ and
Jan~M.~Pawlowski$^{cd}$ \\
\llap{$^a$}Institut f\"ur Theoretische
  Physik, JLU Gie\ss{}en, Heinrich-Buff-Ring
  16, 35392 Gie\ss{}en, Germany\\
\llap{$^b$}Department of Mathematical Physics, National University
  of Ireland Maynooth, Maynooth, County Kildare, Ireland\\
\llap{$^c$}Institut f\"{u}r Theoretische Physik, Universit\"{a}t
  Heidelberg, Philosophenweg 16, 69120 Heidelberg, Germany and\\
\llap{$^d$}ExtreMe Matter Institute EMMI, GSI Helmholtzzentrum
  f\"{u}r Schwerionenforschung mbH, 
  64291 Darmstadt,
  Germany\\
E-Mail: \email{Jan.Luecker@theo.physik.uni-giessen.de}
}

\abstract{We employ the Dyson-Schwinger equations for quark and gluon propagators
	in order to study QCD with $2+1$ flavours at finite temperature and density.
	In a suitable truncation for these equations, we determine the position of the critical end-point
	as well as the deconfinement temperature at all chemical potentials.
	For the latter, the Polyakov-loop potential is obtained from the QCD propagators.
	This is possible for the first time at finite chemical potential, with implications
	for effective models.}

\FullConference{8th International Workshop on Critical Point and Onset of Deconfinement\\
		 March 11-15\\
		 Napa, California, USA}

\begin{document}

\section{Introduction}

The phase diagram of QCD received a lot of attention over the years.
However, many central questions are still open.
In this talk we present results from recent developments of the Dyson-Schwinger (DSE) approach
to hot and dense QCD, see \cite{Fischer:2012vc,Fischer:2013eca}.
The main motivation for using DSEs is the possibility to apply them to the QCD degrees of freedom,
i.e. quarks and gluons, directly.
This is especially interesting for a better understanding of the deconfinement phase transition.
In effective field theories this has to be modelled by including a Polyakov-loop potential,
which we are able to extract from the propagators at all temperatures and densities.
This allows to study the Polyakov loop at finite density.
Additionally we obtain a related order parameter from the dual condensates.

\subsection{Truncation scheme}

\begin{figure}[ht]
\includegraphics[width=0.4\textwidth]{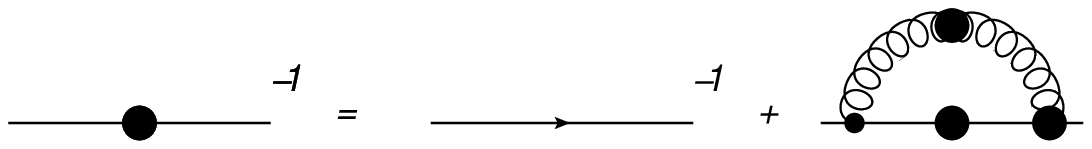}
\hfill
\includegraphics[width=0.55\textwidth]{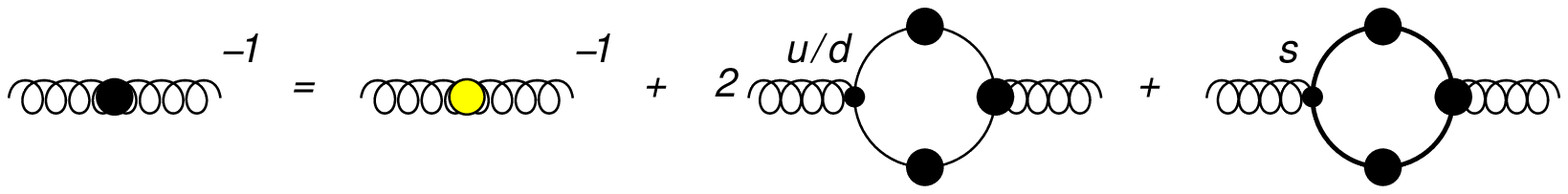}
\caption{The truncated Dyson-Schwinger equations for the quark (left) and gluon (right) propagators.
	The yellow dot in the gluon DSE denotes the quenched propagator.
	\label{fig:DSEs}}
\end{figure}

In the left part of Fig.~(\ref{fig:DSEs}) we show the DSE for the quark propagator.
In order to solve this equation, it is necessary to specify a gluon propagator
and a quark-gluon vertex.
For the vertex we have little information in the medium, and rely on a phenomenological model {\it ansatz},
see \cite{Fischer:2012vc}.
The gluon, on the other hand, can very well be described by a combination of lattice and DSE methods.

Our truncation for the gluon DSE is shown in the right part of Fig.~(\ref{fig:DSEs}).
We use the quenched gluon, with full temperature dependence, from lattice QCD \cite{Fischer:2010fx}.
In our truncation the quenched gluon subsumes the Yang-Mills self-energies in the gluon DSE,
to which we add the quark loop in order to account for unquenching.
In \cite{Fischer:2013eca} we have compared our unquenched gluon results to lattice data
at finite temperature, and found an excellent agreement.
In Fig.~(\ref{fig:DSEs}) we already show the $2+1$ flavour case.
The quark flavours get mixed through the gluon, which is a novel feature of this truncation scheme.

\section{The phase diagram for $2$ and $2+1$ flavours}

\begin{figure}
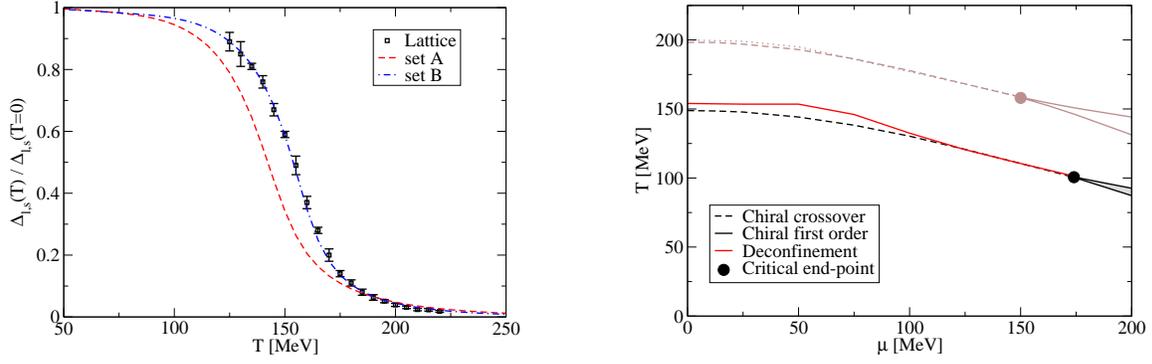

\includegraphics[width=0.45\textwidth]{figures/latticeVSsetsAB2}
\hfill
\includegraphics[width=0.45\textwidth]{figures/phaseDiag}
\caption{The left figure shows the condensate at $\mu=0$ for two parameter sets,
	compared to lattice results from \cite{Borsanyi:2010bp}.
	The right figure shows the resulting phase diagram for $N_f=2$ (upper lines) and $N_f=2+1$ (lower lines).
	Here, $T_c$ is determined from the chiral susceptibility.
	\label{fig:condAndPhaseDiag}}
\end{figure}

We now obtain the quark condensate from a solution of the coupled set of quark and gluon DSEs.
In order to subtract the quadratic divergence present therein, we employ the subtraction
\begin{equation}
\Delta_{l,s} = \langle\bar\psi\psi\rangle_l - \frac{m_l}{m_s}\langle\bar\psi\psi\rangle_s,
\end{equation}
where $l,s$ denote light and strange quarks.
In the left part of Fig.~(\ref{fig:condAndPhaseDiag}), we show the result at $\mu=0$,
where a comparison to lattice data is possible.
In order to test for parameter dependence we employ two parameter sets, labelled A and B.
For set A, we fix the parameters to the pion mass and decay constant in the vacuum.
For set B, we fit to the lattice results of the condensate.
The difference between the sets is rather small, and mainly a shift in $T_c$.

In the right part of Fig.~(\ref{fig:condAndPhaseDiag}), we show the resulting phase diagram
for $N_f=2$ and $N_f=2+1$. There, and for all other results from now on, we use parameter set A.
The deconfinement line is obtained from the dual condensate \cite{Gattringer:2006ci}.
We find the chiral restoration and deconfinement temperatures to be near-by.
The critical end-point is found at $\mu/T > 1$.
The effect of including the strange quark is mainly a reduction of $T_c$, by about $50$ MeV
for all chemical potentials.

\section{Polyakov-loop potential}

Having the quark, gluon and ghost propagators at hand allows us to extract the Polyakov-loop potential,
following the recent development in \cite{Fister:2013bh}.

\begin{figure}[ht]
\centering
\includegraphics[width=0.7\textwidth]{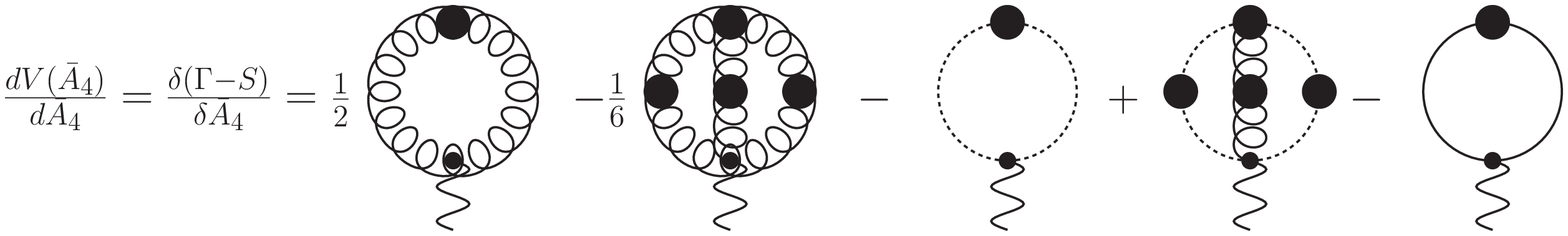}
\caption{The DSE for the background field $\bar{A}_4$.
	\label{fig:DSEbackgroundAcompl}}
\end{figure}

To this end, we solve the DSE for a constant background field $\bar{A}_4$,
which is shown in Fig.~(\ref{fig:DSEbackgroundAcompl}).
This yields the derivative of the Polyakov-loop potential, $V'(\bar{A}_4)$.
Upon integration we gain the potential $V(\bar{A}_4)$, up to an integration constant.

\begin{figure}
\includegraphics[width=0.45\textwidth]{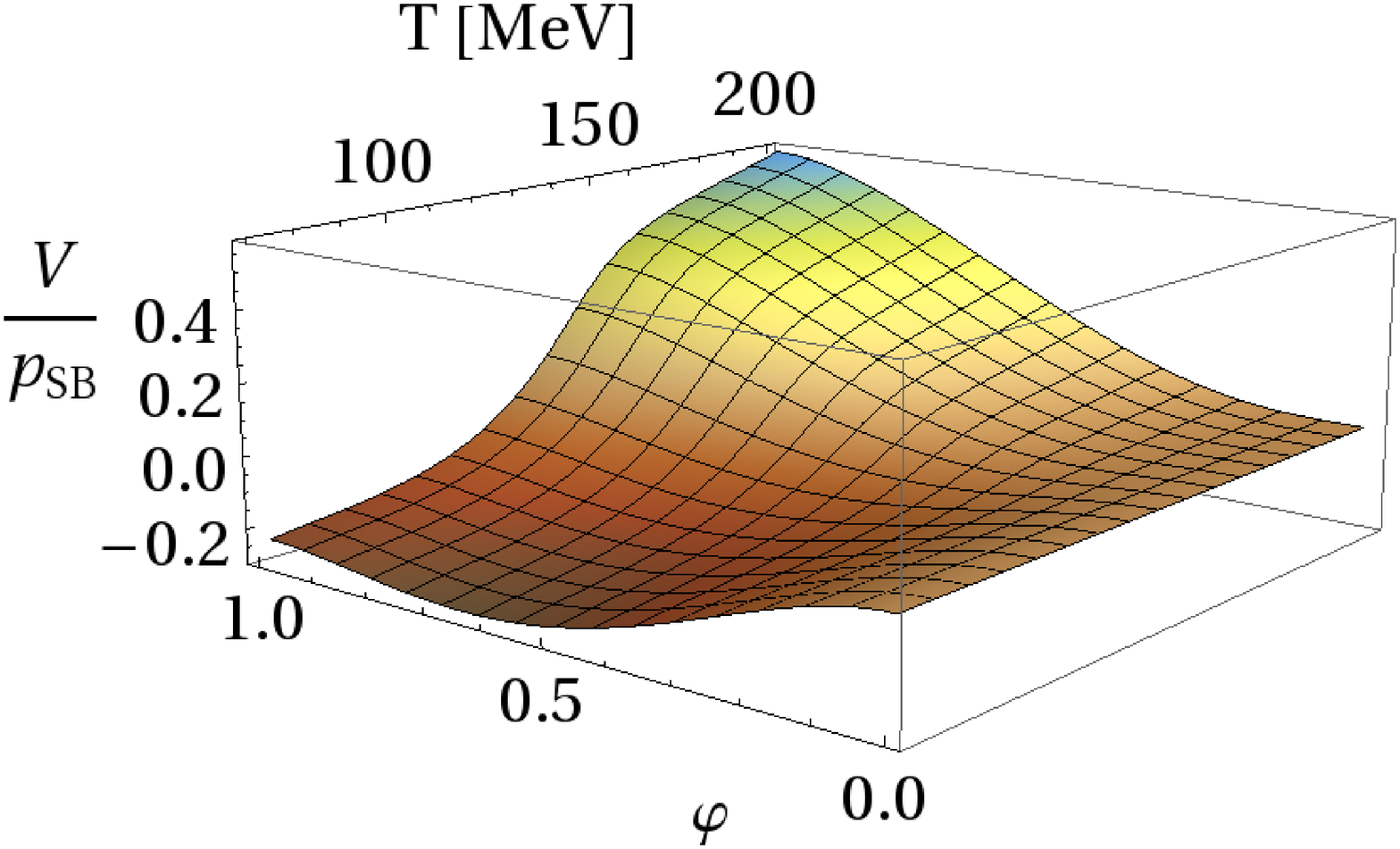}
\hfill
\includegraphics[width=0.45\textwidth]{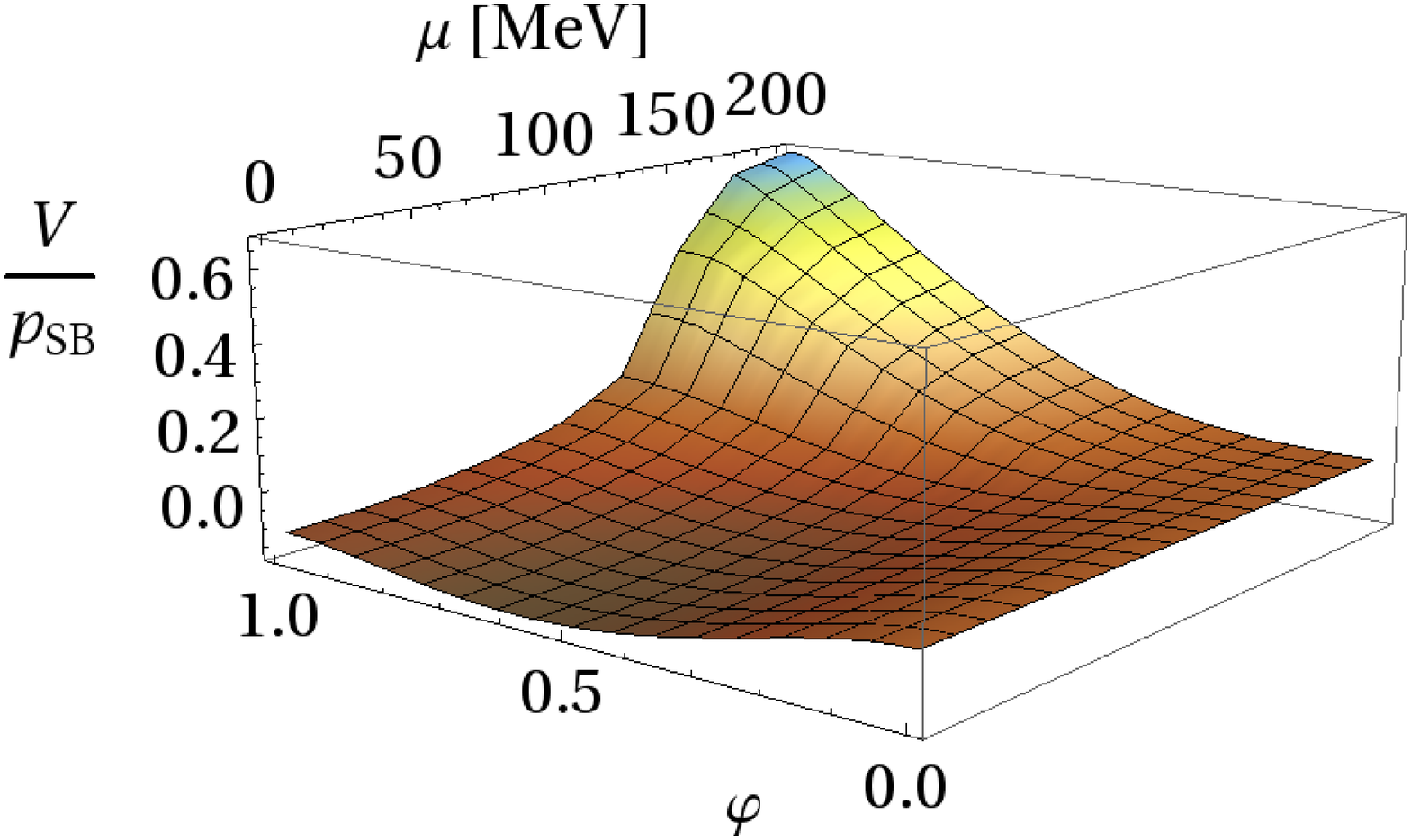}
\caption{The Polyakov-loop potential for $\mu=0$ (left figure) and $T=115$ MeV (right figure).
	\label{fig:PLpotential}}
\end{figure}

The background field enters the propagators via a shift in the Matsubara sums,
\begin{equation}
\omega_n \rightarrow \omega_n +2\pi T \varphi,
\end{equation}
where $\varphi$ is an Eigenvalue of $\bar{A}_4$.
There, we restrict ourselves to the $\lambda^3$ component.
We can then obtain an upper bound for the expectation value of the Polyakov loop,
\begin{equation}
L[\langle A_4 \rangle] = \frac{1+2\cos(\varphi\pi)}{3} \ge \langle L[A_4]\rangle.
\end{equation}
The results for $V(\bar{A}_4)$ at $\mu=0$ as a function of $T$ and at $T=115$ MeV as a function of $\mu$ is shown
in Fig.~(\ref{fig:PLpotential}).
We find a minimum close to the confining value $\varphi = 2/3$ below $T_c$.
Above $T_c$, the minimum moves to smaller $\varphi$, i.e. a larger Polyakov loop.

\begin{figure}
\includegraphics[width=0.45\textwidth]{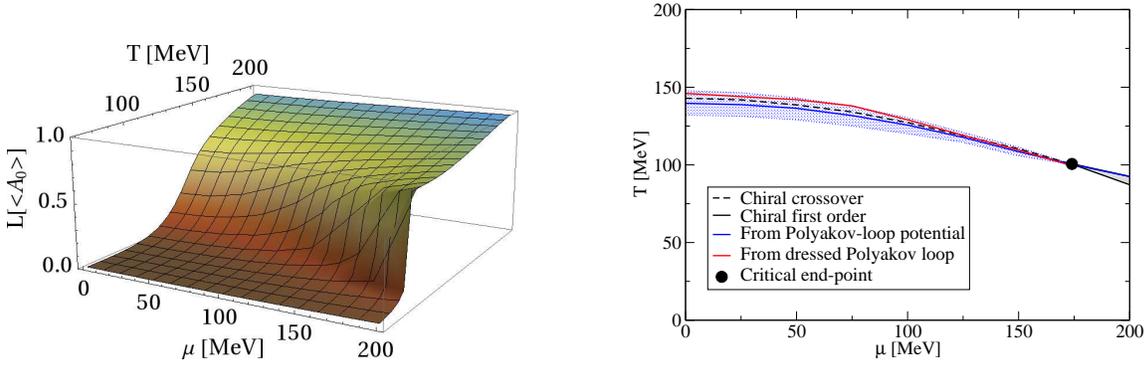}
\hfill
\includegraphics[width=0.45\textwidth]{figures/phaseDiag2ddT}
\caption{The left figure shows the Polyakov loop for all $T$ and $\mu$.
	The right figure shows the resulting phase diagram.
	Here we use $N_f=2+1$, and determine $T_c$ from the inflection point (cf. Fig.~(\protect\ref{fig:condAndPhaseDiag})).
	\label{fig:PLandDiag}}
\end{figure}

We can then go forward, and extract the Polyakov loop of the background field at all $T$ and $\mu$.
The result is shown in the left part of Fig.~(\ref{fig:PLandDiag}).
We clearly find a nearly vanishing value below $T_c$, a rising behaviour around $T_c$
and deconfinement above.
The resulting phase transition line is shown in the right part of Fig.~(\ref{fig:PLandDiag})
together with the phase transition from the quark condensate and the dual condensate.
Clearly all three order parameters show a crossover at similar temperatures.
At the critical end-point all definitions of $T_c$ agree.

To summarise, we presented a truncation for the quark and gluon DSEs which
captures the expected properties of unquenched QCD.
We find a critical end-point, and near-by phase transitions from
the quark condensate, the dual condensate and the Polyakov loop
from the Polyakov-loop potential.

\section{Acknowledgements}

This work has been supported by the Helmholtz Young Investigator Grant VH-NG-332
and the Helmholtz  International Center for FAIR within the LOEWE program of the State of Hesse.

\end{document}